\documentclass{ws-procs9x6}
\usepackage{slashed}
\begin{document}

\title{Octet-baryon masses in finite space}

\author{Xiu-Lei Ren$^{1}$, Lisheng Geng$^{1,2,*}$ and Jie Meng$^{1,2,3,4}$}

\address{
$^1$School of Physics and Nuclear Energy Engineering, Beihang University,\\  Beijing 100191,  China\\
$^2$Research Center for Nuclear Science and Technology, Beihang University,\\ Beijing 100191, China\\
$^3$State Key Laboratory of Nuclear Physics and Technology, School of Physics,
\\Peking University, Beijing 100871, China\\
$^4$Department of Physics, University of Stellenbosch, Stellenbosch, South Africa\\
$^{*}$E-mail: lisheng.geng@buaa.edu.cn
}

\begin{abstract}
We report on a recent study of finite-volume effects on the lowest-lying octet baryon masses using the covariant baryon chiral perturbation theory up to next-to-leading order by analysing the latest $n_f=2+1$ lattice QCD results from the NPLQCD Collaboration.
\end{abstract}


\bodymatter

\section{Introduction}\label{sec1}

Lattice QCD (LQCD) simulations have made remarkable progress in studies of strong-interaction physics (see, e.g., Refs.~\cite{Bazavov:2009bb,Hagler:2009ni,Fodor:2012ll}). Recently, the ground state baryon spectrum
has been calculated with 2+1 flavors and agreement with experimental data up to a few percent has been achived~\cite{Loud:2009lt,Aoki:2009,Lin:2009,Durr:2008,Alexandrou:2009,Aoki:2010,Bietenholz:2010,Beane:2011pc}. However,there are still a few obstacles in present lattice simulations~\cite{Fodor:2012ll}. To obtain physical results, one needs to extrapolate the simulated results to the physical point, i.e., $m_{u/d}\rightarrow m_{u/d}({\rm phys.})$,
 $L(T)\rightarrow\infty$, $a\rightarrow 0$, where $m_{u/d}$ are the masses of $u$ and $d$ quarks, $L(T)$ is the spacial (temporal) lattice size, and $a$ is the lattice spacing.

Chiral perturbation theory (ChPT)~\cite{Weinberg:1978kz,Gasser:1983yg,Gasser:1984gg,Gasser:1987rb,Bernard:1995dp,Pich:1995bw,Bernard:2007zu,Scherer:2009bt}
provides a model-independent way to perform the extrapolation in light quark masses--chiral extrapolation--and to study finite-volume effects. Due to the non-zero baryon mass in the chiral limit, baryon chiral perturbation theory (BChPT) has long suffered from the so-called power-counting-breaking problem~\cite{Gasser:1987rb}. In order to deal with this problem, several recipes have been proposed. The most widely-used are the heavy baryon (HB) ChPT~\cite{Jenkins:1991hb} , the infra-red (IR) BChPT~\cite{Becher:1999ir} and the covariant baryon ChPT with extended-on-mass-shell (EOMS) scheme~\cite{Gegelia:1999eo,Fuchs:2003ms}.

In the 2-flavor space, the finite-volume effects of nucleon masses have been studied using HBChPT and IR BChPT up to next-to-next-to-leading order (NNLO)~\cite{AliKhan:2003cu}. The authors concluded that at NNLO relativistic ChPT can describe well the finite-volume corrections. However, a detailed study of  finite-volume effects using three-flavor BChPT is still missing. In this talk  we report on
a first systemic study of  finite-volume corrections to the masses of ground-state octet baryons using the EOMS BChPT by fitting the NPLQCD~\cite{Beane:2011pc} LQCD data.

\section{Theoretical Framework}
Physically, finite-volume corrections can be easily understood: Because of the existence of space-time boundaries, the allowed momenta of virtual particles become discretized.
In LQCD simulations of zero-temperature physics the temporal extent is generally larger than the spacial extent such that the integral in the temporal dimension can be treated as if it extends from $-\infty$ and $\infty$. As a result, only the integral in the spacial dimensions should be replaced by an infinite sum.

In a finite hypercube, the following differences are defined as the finite-volume corrections:
\begin{equation}
\delta G_N = G_N(L)-G_N(\infty),\quad \delta G_D = G_D(L)-G_D(\infty).
\end{equation}
where $G_{N/D}(L)$ and $G_{N/D}(\infty)$ denote the integrals calculated in a finite hypercube and in infinite space-time. Therefore, the octet-baryon masses at NLO EMOS BChPT in a finite box have the following form (see Ref.~\cite{Geng:2011fvc} for the definitions of couplings $\xi^{(a,b,c)}_{\mathcal{B},\phi}$ and loop functions $H_{B}^{(b,c)}(m_{\phi})$):
\begin{equation}
  \begin{split}
    M_{\mathcal{B}}&= M_0-\sum\limits_{\phi=\pi,K}\xi^{(a)}_{\mathcal{B},\phi}(b_0, b_D, b_F)\cdot m^2_{\phi} \\
    &\quad+ \frac{1}{(4\pi F_0)^2}\sum_{\phi=\pi,K,\eta}
    \left[~\xi_{\mathcal{B},\phi}^{(b)}(D, F)
    \left(H_{B}^{(b)}(m_{\phi})+\delta G_N(L)\right)\right.\\
    &\qquad\qquad \qquad \qquad \qquad \left.+\xi_{\mathcal{B},\phi}^{(c)}(\mathcal{C})
    \left(H_{B}^{(c)}(m_{\phi})
    +\delta G_D(L)\right)~\right].
  \end{split}
\end{equation}

\section{Results and Discussions}

\begin{figure}[t]
\psfig{file=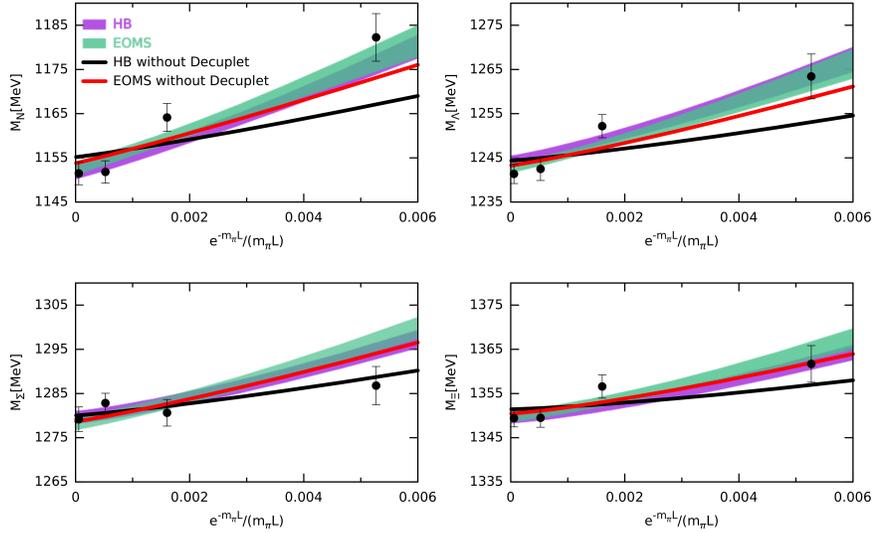,width=4.5in}
\caption{The NPLQCD octet mass data~\cite{Beane:2011pc} fitted with NLO covariant ChPT and HBChPT. The bands are the full results at the 68\% confidence level and the solid (dashed) lines
are the best fits with $C=0$.\label{fig:res}}
\end{figure}

The fitted results using both EOMS BChPT and HBChPT are shown in Fig.~\ref{fig:res}. Both methods provide a reasonable fit to the lattice data with similar quality ($\chi^2$/d.o.f.=1.6). The effects of the virtual decuplet baryons can be best seen by fitting the NPLQCD data with $C=0$. The corresponding results are shown  by the solid (dashed) lines in Fig.~\ref{fig:res}. It is clear that in the fit the octet-decuplet transition plays a larger role in HB than in EOMS BChPT. In fact, in HBChPT virtual decuplet baryons play an even larger role than those of virtual octet baryons, which seems to be a bit unnatural (for a relevant discussion, see, e.g., Ref.~\cite{Geng:2009hh}). In Ref.~\cite{Beane:2011pc}, it was concluded that the decuplet contributions must be taken into account. Our studies show that this is indeed the case, but more so in the HBChPT than in the covariant ChPT.

Using the LECs determined in the fit of the NPLQCD data,
we also performed a chiral extrapolation. 
The EOMS extrapolations are in much better agreement with the experimental masses than the HB extrapolations (see Table III of Ref.~\cite{Geng:2011fvc}
), which is consistent with the finding of Ref.~\cite{MartinCamalich:2010fp}.

\section{Summary and conclusions}
We have studied finite-volume corrections to the octet baryon masses by analyzing the latest $n_f=2+1$ NPLQCD data with EOMS BChPT and with HBChPT. It was shown that although both approaches can describe the lattice data reasonably well, the underlying physics is different: Decuplet contributions play a less important role in EOMS BChPT than in HBChPT at next-to-leading order because relativistic corrections enhance virtual octet contributions and reduce intermediate decuplet contributions.

\section{Acknowledgements}
L. S. Geng acknowledges support from the Fundamental Research Funds for
the Central Universities and the National Natural Science Foundation of
China (Grant No. 11005007).

\end{document}